\documentclass{article}
\usepackage[T1]{fontenc} 
\usepackage[utf8]{inputenc} 
\usepackage{ismir,amsmath,cite,url}
\usepackage{caption}
\usepackage{subcaption}
\usepackage{graphicx}

\usepackage{color}
\usepackage{kotex}

\usepackage[firstpage]{draftwatermark}
\definecolor{lightgray}{rgb}{0.9,0.9,0.9}
\definecolor{darkgray}{rgb}{0.4,0.4,0.4}
\SetWatermarkFontSize{12pt}
\SetWatermarkScale{1.1}
\SetWatermarkAngle{90}
\SetWatermarkHorCenter{202mm}
\SetWatermarkVerCenter{170mm}
\SetWatermarkColor{darkgray}
\SetWatermarkText{Late-Breaking / Demo Session Extended Abstract, ISMIR 2024 Conference}
\def\lbd{}	
\title{Towards Computational Analysis of Pansori Singing}

\threeauthors
  {Sangheon Park} {Georgia Institute of Technology \\ {\tt sangheon@gatech.edu}}
  {Danbinaerin Han} {KAIST \\ {\tt naerin71@kaist.ac.kr}}
  {Dasaem Jeong} {Sogang University \\ {\tt dasaemj@sogang.ac.kr}}

\def\authorname{S. Park, D. Han, and D. Jeong}

\begin{document}

\maketitle

\begin{abstract}

\textit{Pansori} is one of the most representative vocal genres of Korean traditional music, which has an elaborated vocal melody line with strong vibrato. Although the music is transmitted orally without any music notation, transcribing pansori music in Western staff notation has been introduced for several purposes, such as documentation of music, education, or research. 

In this paper, we introduce computational analysis of pansori based on both audio and corresponding transcription, how modern Music Information Retrieval tasks can be used in analyzing traditional music and how it revealed different audio characteristics of what pansori contains.

\end{abstract}
\section{Introduction}\label{sec:introduction}

\textit{Pansori}, a traditional Korean monologue musical storytelling performed by a solo singer with drum accompaniment, is renowned for its emotional depth and diverse musical expressions\cite{pansori}. The music was transmitted orally since the late 17th century without musical notation, and each singer made their own improvisation and variation, which enriched the repertoire and expression of pansori. 

Musicological research on pansori often relies on manual transcription of music in Western staff notation~\cite{Jeokseongga, Simcheongga, Wang2024}. 
While using Western notation for transcribing pansori is necessary for analyzing scales and rhythmic patterns, allowing for quantification and comparison, it has clear limitations in capturing the flexible melodies and improvisational nature of the performance. 

To address the limitations of relying solely on Western notation, we incorporate a computational approach by aligning the audio recordings of pansori with their corresponding transcriptions. This approach allows us to directly examine the nuances of vocal performance, including variations in pitch, timing, and ornamentation that are difficult to capture through notation alone.
By analyzing the audio recordings of a renowned pansori musician alongside her own transcriptions, we gain fuller insights into the performance practices and interpretive decisions that shape pansori’s rich and expressive musical landscape.

\section{Dataset}
Among five classic pieces of pansori, we focused on \textit{Jeokbyeokga} (적벽가, The Song of the Red Cliff), which is based on the famous Battle of Red Cliffs from Romance of the Three Kingdoms (三國演義). 
We used a recording sung by Chae Soojung (채수정), which was released in 2024~\cite{chae2024}.  She also published a transcription of entire \textit{Jeokbyeokga} sung by her mentor, Park Songhee (박송희), in Western staff notation. As Chae strictly followed her mentor's version in most of the parts~\cite{chae2024}, this transcription and her own recording make a pair that is valuable to analyze. 

Among 30 different \textit{daemok} of Jeokbyeokga, we selected 8 of them using \textit{Joongmori} \textit{jangdan}. \textit{Daemok} is a section of the song in pansori. \textit{Jangdan} is a metrical concept in Korean traditional music, which includes both tempo and rhythmic pattern and \textit{Joongmori} is one type of jangdan in mid-fast tempo and typically notated in 12/4 meter in western notation~\cite{min2014great}. The total length of eight daemok was 166 measures in 12/4 meter. 

\begin{figure}[t]
    \centering
    \includegraphics[width=0.5\textwidth]{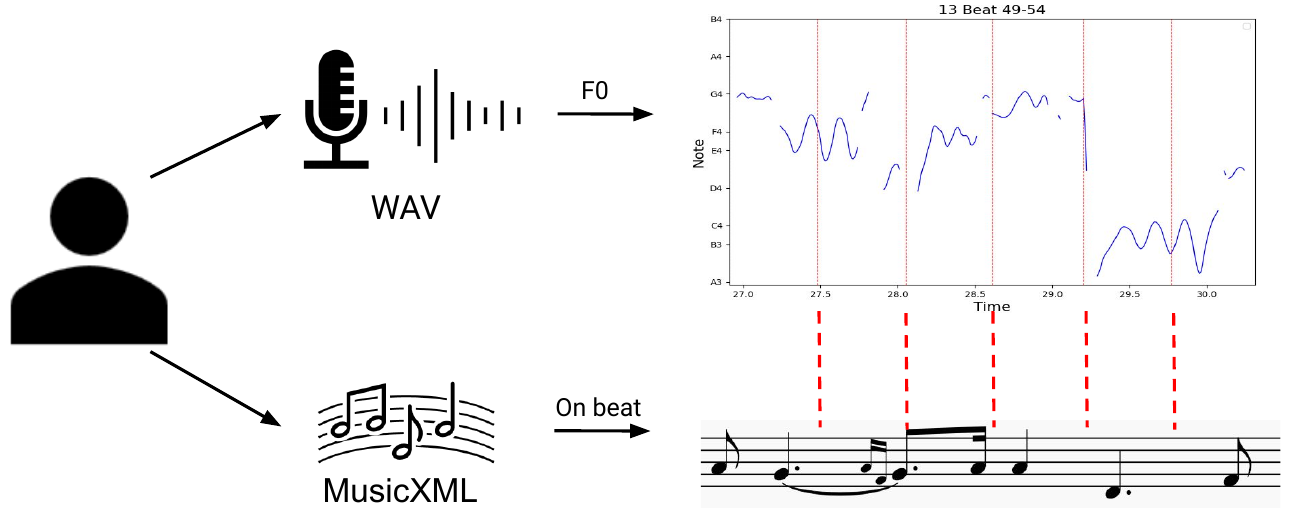}
    \caption{Overview of data visualization pipeline from the pansori singer to data.}
    \label{fig:pipeline}
\end{figure}

\section{Method}
To extract the fundamental frequency (F0) contour, which captures the most essential melodic information from each audio recording, we employed the CREPE algorithm~\cite{kim2018crepe}. 
% CREPE provides F0 estimates at 10 ms intervals, each accompanied by a confidence score ranging from 0 to 1. 
To reduce the noise of the extracted F0, we filtered out F0 values with confidence scores below 0.6. 
Additionally, to avoid including pitch values outside the typical range of a singer's voice, we excluded frequencies below 350 Hz and above 1000 Hz.

For beat detection, we utilized the \texttt{madmom} library~\cite{bock2016madmom}, specifically its recurrent neural network beat detection tool. We then manually annotated the beats to ensure that each measure contained exactly 12 beats, maintaining consistency throughout the analysis. Although all daemok were sung in the same joongmori jangdan, in instances where the singer performed a measure with a dynamic tempo and rhythmic complexity, the beat detection did not function as accurately, occasionally resulting in more than 12 beats per measure and inconsistency. For example, in Daemok 11-3 ---a soldiers' sorrowful longing for his wife---, there is a section where the singer performs with an off beat tempo with syncopation. In this case, we had to adjust most of the beat notation manually. However, in daemok with rhythmic consistency and even interval between beats, the madmom beat detection tool generally demonstrated stable performance.

\subsection{Pitch Histogram}

\begin{figure}[t]
    \centering
    \includegraphics[width=0.5\textwidth]{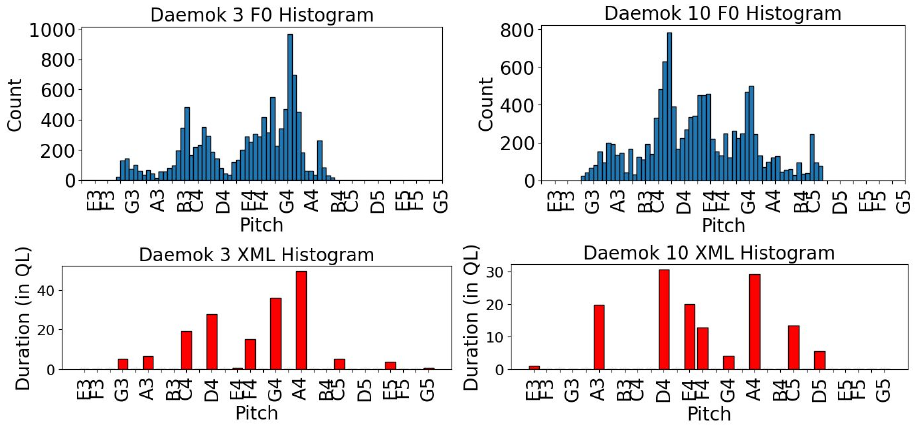}
    \caption{Pitch histograms of two daemok, No.~3 and No.~10, which have different modes, Ujo and Gyemyeonjo. Histogram shows F0 contour counts, and note durations from MusicXML transcriptions.}
    \label{fig:histogram}
\end{figure}
To analyze and verify the characteristics of the vocal performance, we generated two types of histograms: one representing the distribution of F0 values and the other capturing note duration extracted from the MusicXML file.

\subsection{Pattern Detection}
To analyze note patterns that frequently appear throughout different daemok, we converted all the notes from the MusicXML file into a word-like encoding. Each note was converted as a single text word that contains note pitch value and note duration. Then, we grouped these patterns using n-gram algorithm, especially 2, 3, 4, and 6 grams. From all sorted n-gram patterns that we have, we exported the corresponding F0 value of each pattern and plotted them by pattern so that we can compare the F0 value of the pattern in different parts of the daemok or even in different daemok. 

\section{Results}

\subsection{Pitch Histogram}
Analyzing the mode of each daemok in Pansori is key in understanding the structure and musical language of genre. \textit{Jeokbyeokga}, the piece that we used for data employs two modes: \textit{Gyemyeonjo}, and \textit{Ujo}. A significant characteristic of Pansori is the modulation between modes within a single piece. By examining the pitch content and ornamentation used in different sections of the daemok, we can identify the dominant mode in each part and can evaluate which mode dominates throughout the piece.

A pitch histogram serves as an effective tool for visualizing the note distribution in the daemok. By comparing the pitch occurrences with the basic Ujo (D-F-G-A-C) and key notes in Gyemyeonjo(D-FE-(G)-A-C), we can observe dominant mode in each section. Figure~\ref{fig:histogram} presents the example from Daemok 3, which is about the episode of \textit{The Three Visits to the Thatched Cottage}, and Daemok 10, which describes soldiers playing and crying. In Gyemyeonjo, F-E is frequently used to depict crying sounds or sad emotions. The histogram shows a greater use of Gyemyeonjo with frequent use of E note in Daemok 10, both in F0 and the transcription. 
 
\begin{figure}[t]
    \centering
    \begin{subfigure}{\columnwidth}
        \centering
        \includegraphics[width=\linewidth]{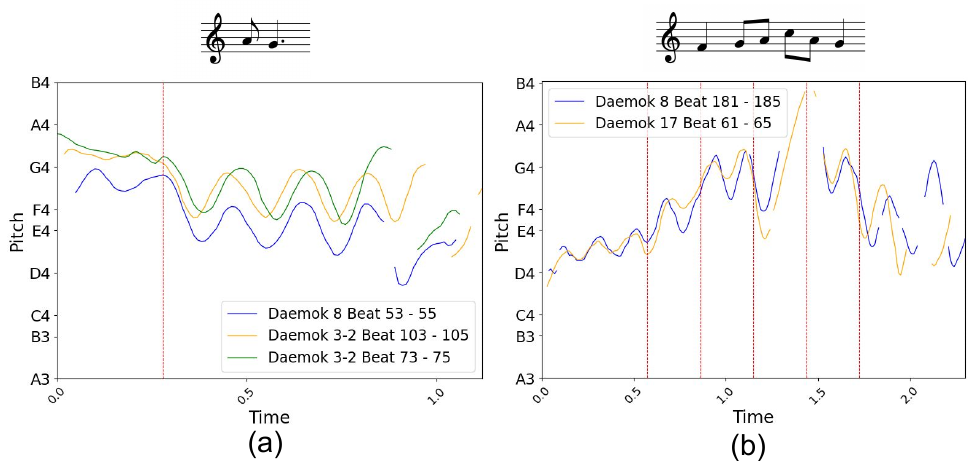}
        % \label{fig:plot1a}
    \end{subfigure}
    \begin{subfigure}{\columnwidth}
        \centering
        \includegraphics[width=\linewidth]{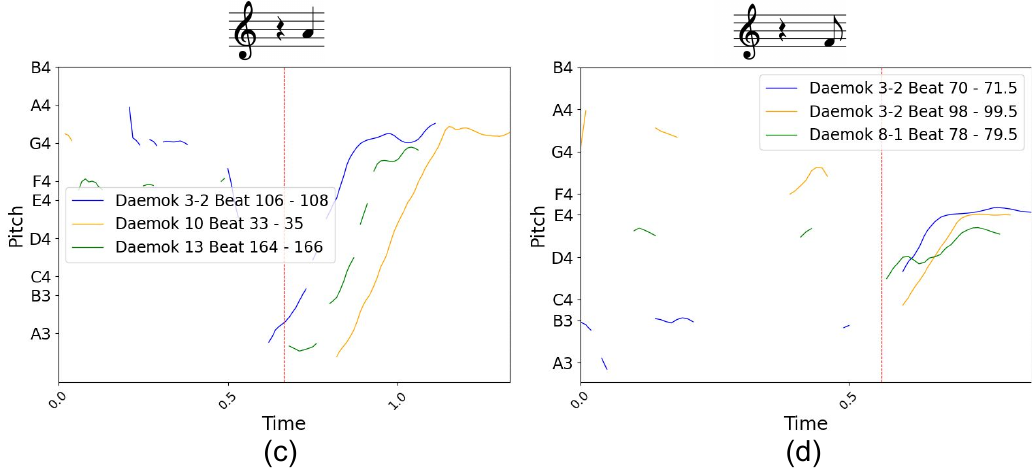}
        % \label{fig:plot1c}
    \end{subfigure}
    \caption{F0 contours of selected patterns}
    \label{fig:plot1}

\end{figure}

\subsection{Pattern Analysis Based n-gram}

One of the major characteristics of pansori is its dynamic vibrato, which plays a significant role in conveying emotion and enhancing the musicality of the performance. As illustrated in Figure~\ref{fig:plot1}~(a), within the common rhythmic pattern of joongmori, vibrato across different daemok exhibit similarities in both depth and rate. 

Using the n-gram pattern, we could find an example where idiomatic patterns appear in different daemok, as presented in Figure~\ref{fig:plot1}~(b). The F0 contours also show a close overlap, which emphasizes that this pattern was sung almost identically in different daemok. Figure~\ref{fig:plot1}~(c) and (d) shows that the singer sometimes used an ascending portamento when singing a long note after a rest. 

\section{Future Work}
Our future work aims to expand the analysis to include more singers and diverse musical samples. By analyzing the musical styles and patterns transmitted through various schools and traditions, we seek to gain a deeper understanding of the uniqueness of pansori that distinguishes it from other musical forms.

\section{Acknowledgement}
This work was supported by the Ministry of Education of the Republic of Korea and the National Research Foundation of Korea (NRF-2024S1A5C3A03046168).

\bibliography{reference}

\end{document}